\documentclass[a4paper,11pt]{article}
\usepackage{pos}

\title{	Sensitivities to feebly interacting particles: public and unified calculations}

\author*[a,b]{Maksym Ovchynnikov}
\author[c]{Jean-Loup Tastet}
\author[b]{Oleksii Mikulenko}
\author[d,e,f]{Kyrylo Bondarenko}

\affiliation[a]{Institut für Astroteilchen Physik, Karlsruher Institut für Technologie (KIT), Hermann-von-Helmholtz-Platz 1, 76344 Eggenstein-Leopoldshafen, Germany}

\affiliation[b]{Instituut-Lorentz, Leiden University, Niels Bohrweg 2, 2333 CA Leiden, The Netherlands} 
\affiliation[c]{Departamento de Física Teórica and Instituto de Física Teórica UAM/CSIC, Universidad Autónoma de Madrid, Cantoblanco, 28049, Madrid, Spain}
\affiliation[d]{IFPU, Institute for Fundamental Physics of the Universe, via Beirut 2, I-34014 Trieste, Italy}
\affiliation[e]{SISSA, via Bonomea 265, I-34132 Trieste, Italy}
\affiliation[f]{INFN, Sezione di Trieste, SISSA, Via Bonomea 265, 34136, Trieste, Italy}

\emailAdd{maksym.ovchynnikov@cern.ch}
\emailAdd{jean-loup.tastet@uam.es}
\emailAdd{mikulenko@lorentz.leidenuniv.nl}
\emailAdd{kyrylo.bondarenko@sissa.it}

\abstract{The increasing interest in Long-Lived Particles (LLPs) has led to numerous proposed experiments in order to search for them. However, the sensitivity estimates published by these experiments tend to rely on disparate assumptions. To ensure an accurate comparison of their potential to find LLPs, a unified estimation of their sensitivity is therefore required. In this contribution, we introduce \texttt{SensCalc}, a \texttt{Mathematica}-based code that uses a semi-analytic approach to calculate the event rate of GeV-scale LLPs, and we present several case studies.}

\FullConference{%
  The European Physical Society Conference on High Energy Physics (EPS-HEP2023)\\
  21-25 August 2023\\
  In-person
}

\begin{document}
\maketitle

\section{The need for a unified public sensitivity evaluator}
Recently, a lot of experiments have been proposed to search for LLPs in the GeV-scale range~\cite{Antel:2023hkf}. Most of them are at an early stage of research and development, meaning in particular that the setup is constantly changing. To understand the relative potential of each experiment to explore LLPs, the Physics Beyond Colliders (PBC) working group proposed to compare their sensitivities in the LLP parameter space for several benchmark models of Higgs-like scalars, axion-like particles, heavy neutral leptons, dark photons, millicharged particles, and light dark matter~\cite{Beacham:2019nyx}.

This myriad of experiments with changing designs, combined with different descriptions of the LLP phenomenology existing in the literature, have made it particularly difficult to conduct a fair and accurate comparison. Notable difficulties include the different assumptions made about the LLP phenomenology in the calculations done by different experiments, the slow updates of sensitivities when changing the setup, and, last but not least, the lack of access to the tools used to perform the sensitivity calculations by the broader community (outside the collaborations), which affects the transparency of the calculations. For some LLPs, the calculations have been performed by groups outside the collaborations, making the picture even more inhomogeneous.

This lack of consistency poses the need for a public sensitivity evaluator that can calculate the event rates for LLPs at different experiments in a unified way. The requirements for such an evaluator are: i) a good accuracy, comparable with the Monte-Carlo evaluators, ii) its relative simplicity, allowing people to use it without any efforts for already-implemented LLPs and experiments, and also implement new ones easily, and iii) a short runtime when calculating the sensitivity from scratch.

\section{\texttt{SensCalc}}

Semi-analytic calculations~\cite{Bondarenko:2019yob} provide a good starting point for such a unified approach. The idea is to represent the number of events as the integral of the product of several pre-computed quantities: the LLP angle-energy distribution, the geometric acceptance of the experiment, the LLP decay probability, and the acceptance for the decay products. This method has been successfully used for a number of experiments and LLPs~\cite{Boiarska:2021yho,Boyarsky:2022epg,Ovchynnikov:2022its} (see also~\cite{Coloma:2023adi,Batell:2023mdn}).

In ref.~\cite{Ovchynnikov:2023cry}, we have systematically implemented this approach by developing the \texttt{Mathematica}-based code \texttt{SensCalc}.\footnote{Available at \url{https://doi.org/10.5281/zenodo.7957784} and also at \url{https://github.com/maksymovchynnikov/SensCalc}\,.} It is made of four modules: one calculating the tabulated geometric acceptance of the experiment and the decay products' acceptance, one calculating the LLP angle-energy distribution function, one that combines these quantities to compute the tabulated number of events and the sensitivity; and finally one which produces plots.

\begin{figure}[t!]
    \centering
    \includegraphics[width=0.5\textwidth]{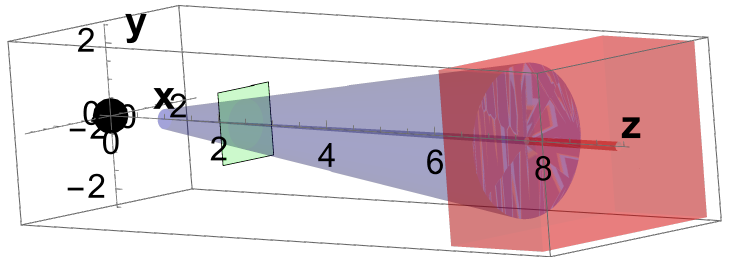}~\includegraphics[width=0.5\textwidth]{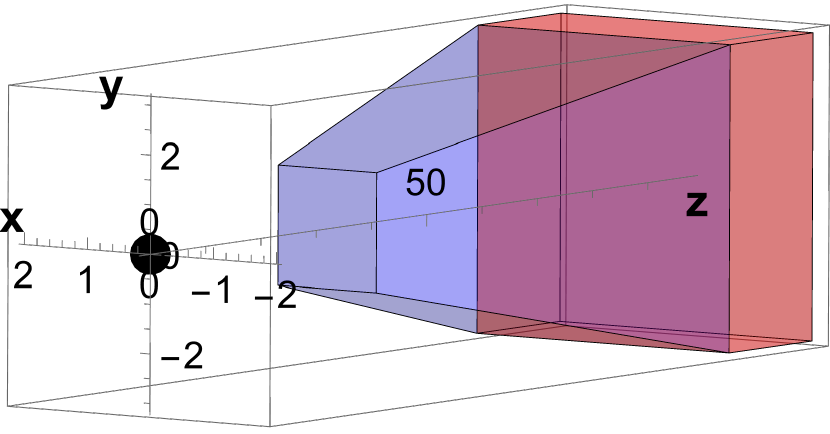}\\\includegraphics[width=0.5\textwidth]{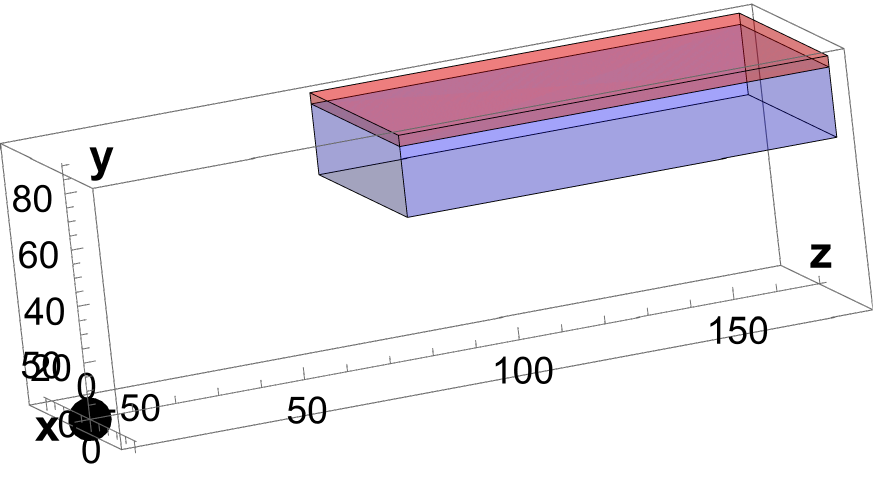}~\includegraphics[width=0.5\textwidth]{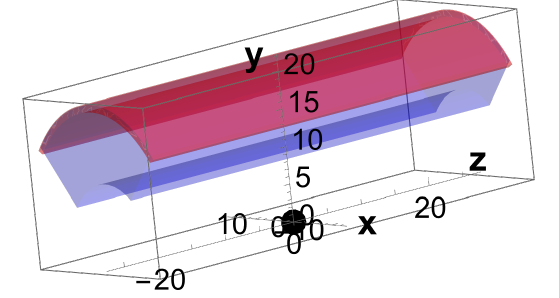}
    \caption{Some of the implemented experiments: the \texttt{Downstream} setup at LHCb~\cite{LHCbdownstream} (\textbf{top left}), SHiP~\cite{Aberle:2839677} (\textbf{top right}), MATHUSLA~\cite{MATHUSLA:2022sze} (\textbf{bottom left}) and the ceiling configuration of ANUBIS~\cite{Satterthwaite:2839063} (\textbf{bottom right}). The black dot shows the location of proton-proton or proton-target interaction; the blue domains indicate the decay volumes and the red domains are detectors.}
    \label{fig:implemented-geometries}
\end{figure}

The implemented LLPs includes heavy neutral leptons with various mixing patterns, Higgs-like scalars with mixing and trilinear coupling to the Higgs boson, dark photons, ALPs coupled to fermions, gluons, or photons, and anomaly-free $U(1)$ mediators coupled to the baryon and lepton currents. Many other models will be implemented in the future. The included experiments are those currently running or proposed at Fermilab and at the SPS, LHC, and FCC-hh (see Fig.~\ref{fig:implemented-geometries} for some examples).

\texttt{SensCalc} features a lot of options, including: specifying the visible LLP decay channels; imposing pre-selection criteria for the decay products, e.g., on the invariant mass of particle pairs, their energy, the spatial separation between tracks, and the impact parameter; estimating the effect of the detector's dipole magnet on the trajectories of the decay products; studying explicitly the impact of various factors entering the number of events (such as the geometric acceptance and decay probability) on the event rate; producing the event rate distribution in terms of the LLP's angle and energy; producing the event density in the plane of the LLP's mass and coupling. \texttt{SensCalc}'s main limitation comes from the fact that it does not produce a detailed event record, so it is not suited for a detailed event analysis or further processing with the full detector simulation. An advanced version of \texttt{SensCalc} that produces the event data is in preparation. 

\texttt{SensCalc} has been cross-checked for various LLPs and geometries against several Monte-Carlo codes (more details can be found in ref.~\cite{Ovchynnikov:2023cry}).

\section{Case studies}
Let us illustrate some interesting applications of \texttt{SensCalc}.

\subsection{Qualitative comparison of the sensitivities of different experiments}
Let us consider a few experiments that differ significantly in their housing facility, geometric placement, and detector equipment. As an example, we will use the SHiP, FASER2, and MATHUSLA experiments, which would be located at the SPS (SHiP) and the LHC (MATHUSLA, FASER2), on-axis (SHiP, FASER2) and off-axis (MATHUSLA). In addition, SHiP and FASER2 are equipped with electromagnetic and hadronic calorimeters, unlike MATHUSLA, which lacks those. The goal is to understand qualitatively the impact of these factors on the sensitivity. 

\begin{figure}[t!]
    \centering
    \includegraphics[width=0.45\textwidth]{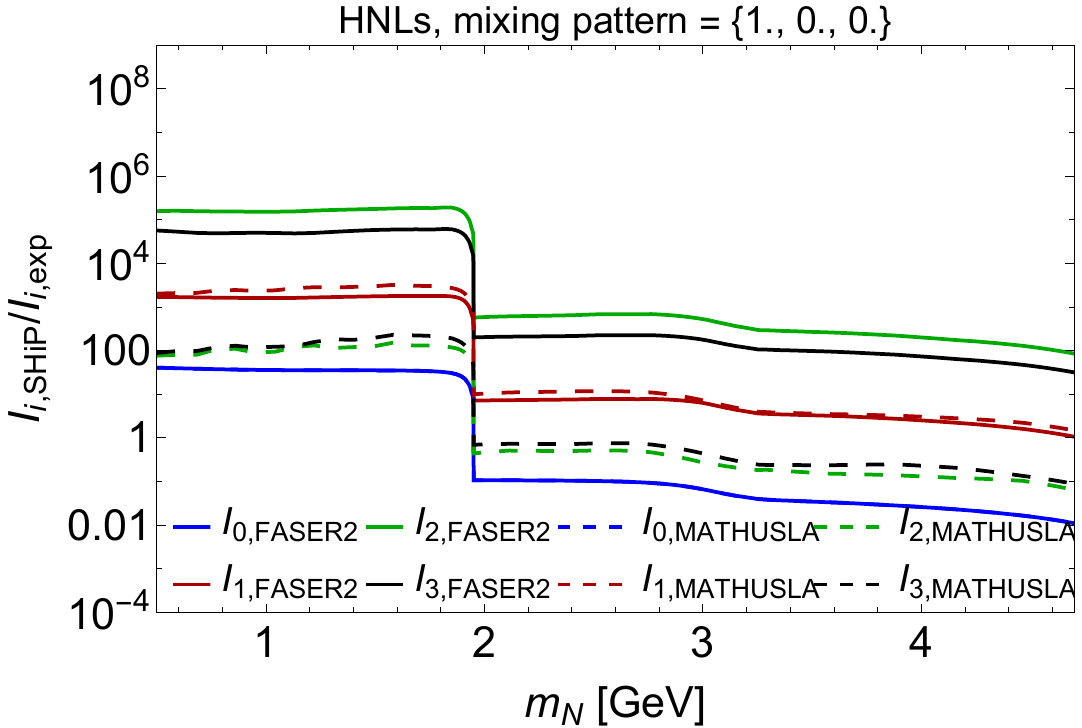}~\includegraphics[width=0.45\textwidth]{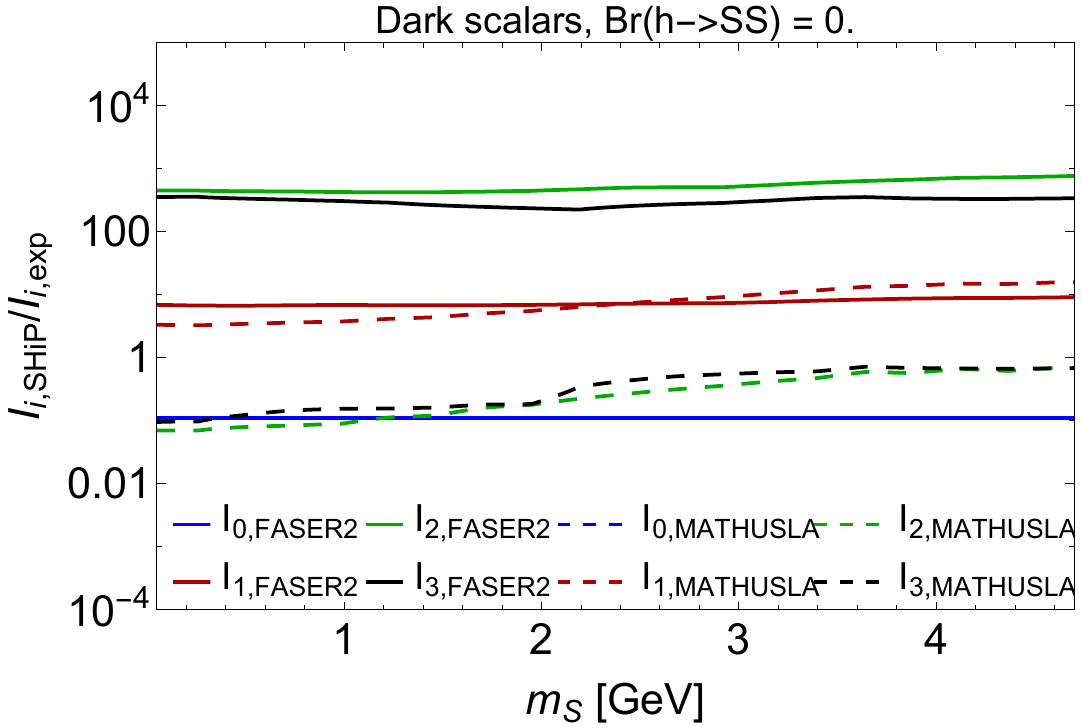}
    \caption{The \emph{ratio} of the $\mathcal{I}_{i}$ quantities (see text for details) for SHiP$/$FASER2 (solid lines) or SHiP$/$MATHUSLA (dashed lines) for Heavy Neutral Leptons coupled solely to the electron neutrino (\textbf{left}) and Higgs-like scalars mixing with the Higgs boson (\textbf{right}). Note that the blue lines overlap because of the same luminosity to be collected during the runtime of FASER2 and MATHUSLA.}
    \label{fig:events-ratio}
\end{figure}

For this purpose, we will consider the regime of large LLP lifetimes, such that the decay probability is in the linear regime for all of the mentioned experiments, 
\begin{equation}
P_{\text{decay}}\approx \frac{\Delta L_{\text{decay volume}}}{c\tau_{\text{LLP}}} m_{\text{LLP}}\langle p_{\text{LLP}}^{-1}\rangle
\label{eq:dec-prob}
\end{equation}
and study the mass dependence of the total number of produced LLPs (denoted by $\mathcal{I}_{0}$), the fraction of LLPs intersecting the decay volume ($\mathcal{I}_{1}$), the fraction of those decaying inside ($\mathcal{I}_{2}$), and the subset of these events for which the decay products meet the decay acceptance criteria ($\mathcal{I}_{3}$), which corresponds to the total number of events (more details on the procedure may be found in Sec.~IV.A of ref.~\cite{Ovchynnikov:2023cry}). The ratios of these quantities between the three considered experiments, for the models of Heavy Neutral Leptons and Higgs-like scalars, are shown in Fig.~\ref{fig:events-ratio}.

From the figure, we see that the total number of produced LLPs (blue lines) from $B$ mesons (above $2\,\mathrm{GeV}$ for HNLs, everywhere for Higgs-like scalars) is larger at the LHC, whereas the amount of the LLPs produced from $D$ mesons (below $2\,\mathrm{GeV}$ for HNLs) is much larger at the SPS. The situation changes once we require (red lines) the LLPs to pass through the decay volumes of SHiP, FASER2, and MATHUSLA, as FASER2 has a much smaller geometric acceptance than SHiP, while MATHUSLA is located off-axis where the flux of $B$ is suppressed. Next, taking also into account the decay probability~\eqref{eq:dec-prob} (green lines), we see that the ratio SHiP$/$FASER2 further increases as SHiP has a much longer decay volume and LLPs at the SPS have a much larger $\langle p_{\text{LLP}}^{-1}\rangle$, while SHiP$/$MATHUSLA decreases, since $\langle p_{\text{LLP}}^{-1}\rangle$ at MATHUSLA is much larger due to the off-axis placement. Considering finally the decay acceptance (black lines), the ratio SHiP$/$FASER2 decreases (since decay products at FASER2 always point to the detector), while SHiP$/$MATHUSLA increases (as decay products at MATHUSLA are low-energetic and may easily miss the detector).

\subsection{ALPs coupled to fermions and Higgs-like scalars}

Two examples of long-lived particles for which the published sensitivities~\cite{Beacham:2019nyx,Antel:2023hkf} have been computed under disparate assumptions are ALPs coupled to fermions (denoted by BC10) and Higgs-like scalars (BC4). In particular, some of experiments use the inclusive description of the dark scalar production from $B$ mesons (which may underestimate the number of events by a factor of 10 or higher in the domain of large masses) and are missing production and decay channels for the ALPs coupled to fermions (the interested reader is referred to refs.~\cite{Ovchynnikov:2023cry,DallaValleGarcia:2023xhh}). Finally, some of the setups used to compute the sensitivity are already outdated, or even disagree with all published descriptions of the experiment.

\begin{figure}[t!]
    \centering
    \includegraphics[width=0.45\textwidth]{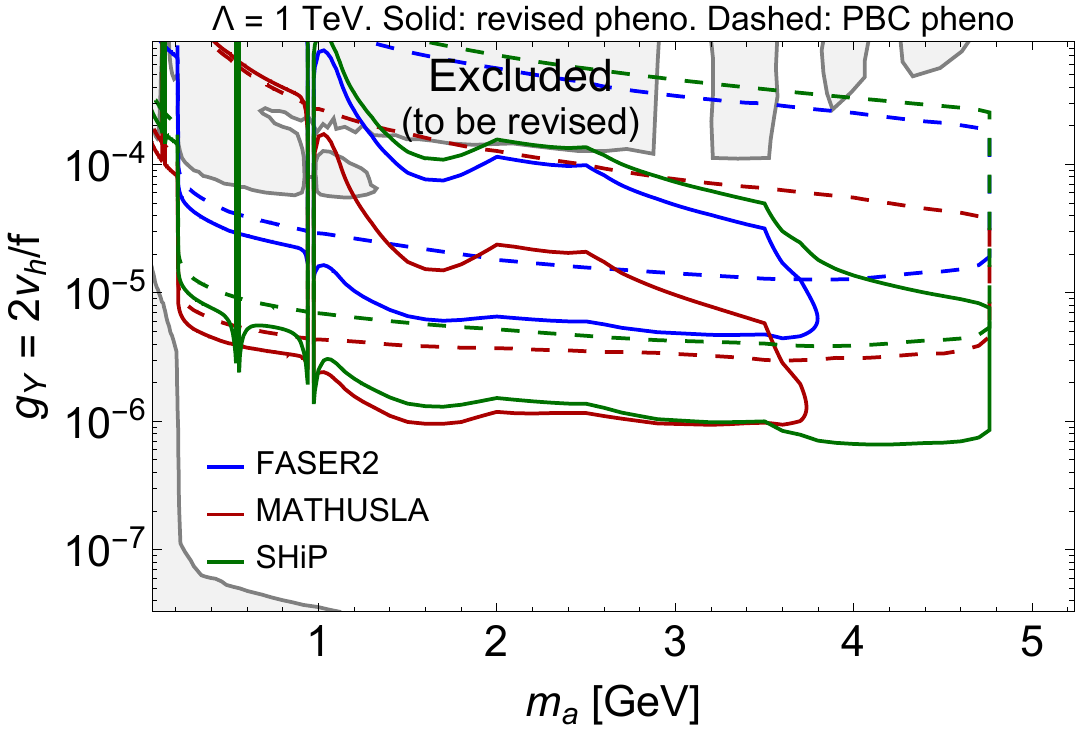}~\includegraphics[width=0.45\textwidth]{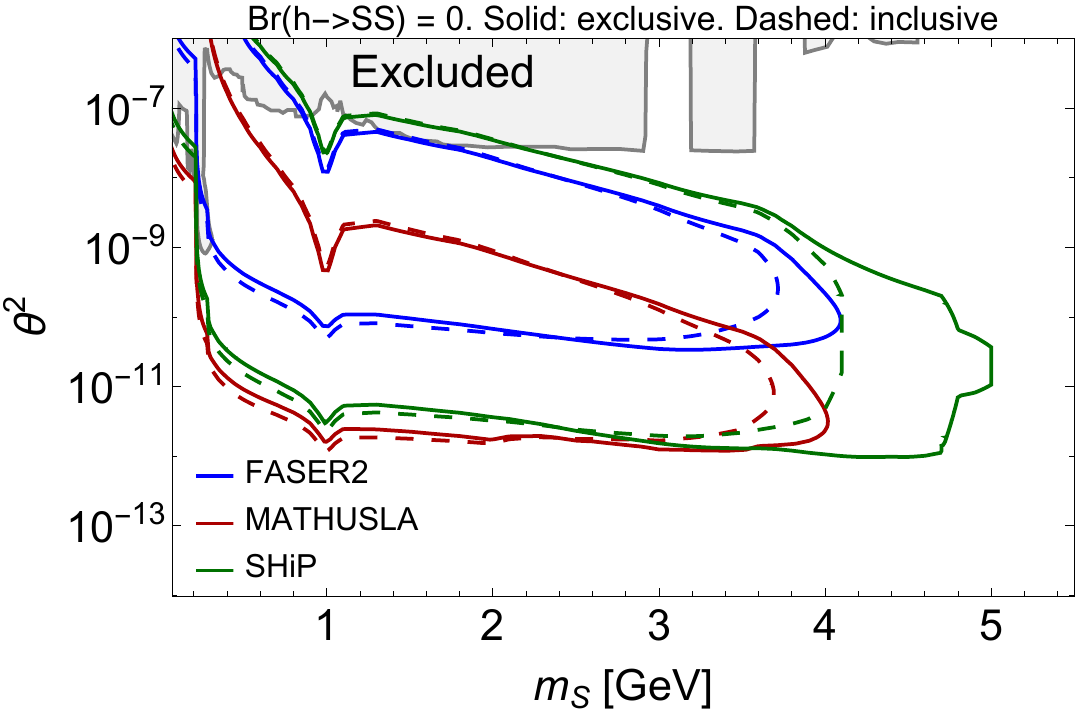}
    \caption{Sensitivities of the SHiP, FASER2, and MATHUSLA experiments to ALPs that couple universally to fermions (\textbf{left}) and to dark scalars mixing with the Higgs boson (\textbf{right}). The solid lines denote the phenomenology newly derived in refs.~\cite{DallaValleGarcia:2023xhh,Boiarska:2019jym}, while the dashed lines assume the inclusive production of dark scalars and, for ALPs, the absence of hadronic decays and the production only through the channels $B\to K+a$ and $B\to K^{*}+a$, similarly to what was previously used in ref.~\cite{Beacham:2019nyx}. For SHiP, we adopted the decay products selection criteria specified in ref.~\cite{Aberle:2839677} and the setup from ref.~\cite{Ahdida:2867743}. For FASER2 and MATHUSLA, we did not assume any selection on the decay products beyond the geometric acceptance itself, and we used the setups from ref.~\cite{MATHUSLA:2022sze,FASER:2018bac}.}
    \label{fig:sensitivity-scalar-alp}
\end{figure}

In Fig.~\ref{fig:sensitivity-scalar-alp}, we show the sensitivities of the SHiP, MATHUSLA, and FASER2 experiments to the scalars and ALPs under different sets of assumptions about their phenomenology, in order to demonstrate the impact of assumptions on the parameter space probed. We notably observe that the boundary of the probed region may change by orders of magnitude.

\bigskip
In conclusion, we have presented a public and unified sensitivity evaluator --- \texttt{SensCalc} --- that performs sensitivity calculations in a transparent and robust way and is made available to the whole scientific community. Its aim is to qualitatively improve the exploration of the FIP parameter space, by ensuring greater consistency between the sensitivity studies.

\bibliographystyle{JHEP}
\bibliography{references}

\end{document}